\documentclass[letterpaper]{article}
\usepackage{graphicx}
\usepackage{hyperref}
\hypersetup{
    colorlinks=blue,
    linkcolor=blue,
    filecolor=blue,      
    urlcolor=blue,
    }
\usepackage{geometry}
\usepackage{algorithm}
\usepackage{algpseudocode}

\usepackage{caption}
\usepackage{amsmath}

\usepackage[longnamesfirst,sort]{natbib}
\bibpunct[, ]{(}{)}{;}{a}{,}{,}

\title{Revising price coordination in the classical and neoclassical economics based on elementary cellular automata}

\author{Igor Lugo\thanks{Corresponding author: \href{igorlugo@crim.unam.mx}{igorlugo@crim.unam.mx}}\\Universidad Nacional Autonoma de Mexico (UNAM)\\Centro Regional de Investigaciones Multidisciplinarias (CRIM)\vspace{0.5cm}\\ Martha G. Alatriste-Contreras\\Universidad Nacional Autonoma de Mexico (UNAM)\\ Facultad de Economia \vspace{0.5cm}}

\date{Jun 29, 2026}

\begin{document}

\maketitle

\begin{abstract}
The coordination of prices in economics is one of the most complex phenomena. In particular, the classical and neoclassical approaches related to the economic theory provide some insights into such a complex coordination based on different formulations. However, these formulations have not been successful for explaining simple mechanisms to understand and predict a set of prices that theoretically clears all markets. Consequently, elementary cellular automata can contribute to clarify such a coordination problem by using simple computational rules to describe the theoretical bases of the classical and neoclassical economics. Therefore, we propose to use this type of cellular automata for explaining different escenarios of price coordination in which simple rules of price interactions generate stable and unstable patterns of coordination. We used an explorative data analysis based on the Shannon entropy for computing the uncertainty related to such generated patterns of coordination, and a Monte Carlo simulation approximation based on a Spearman correlation for evaluating the statistical significance of such price coordination.  
Findings suggested that the classical economics provides a consistent approach for understanding the coordination of prices because it emphasizes human interactions based on logical choices related to an objective data. 
On the other hand, the neoclassical approach does not propose any type of mechanism for describing the price coordination. The neoclassical individual is just a spectator and receiver of the unpredictable and supposed event of price coordination. 
As a result, by modeling the economic theory based on computational concepts, we reveal facts and believes behind the classical and neoclassical economics.

\end{abstract}

\section{Introduction}
\begin{figure}[ht]
\centering
\includegraphics[width=12.5 cm]{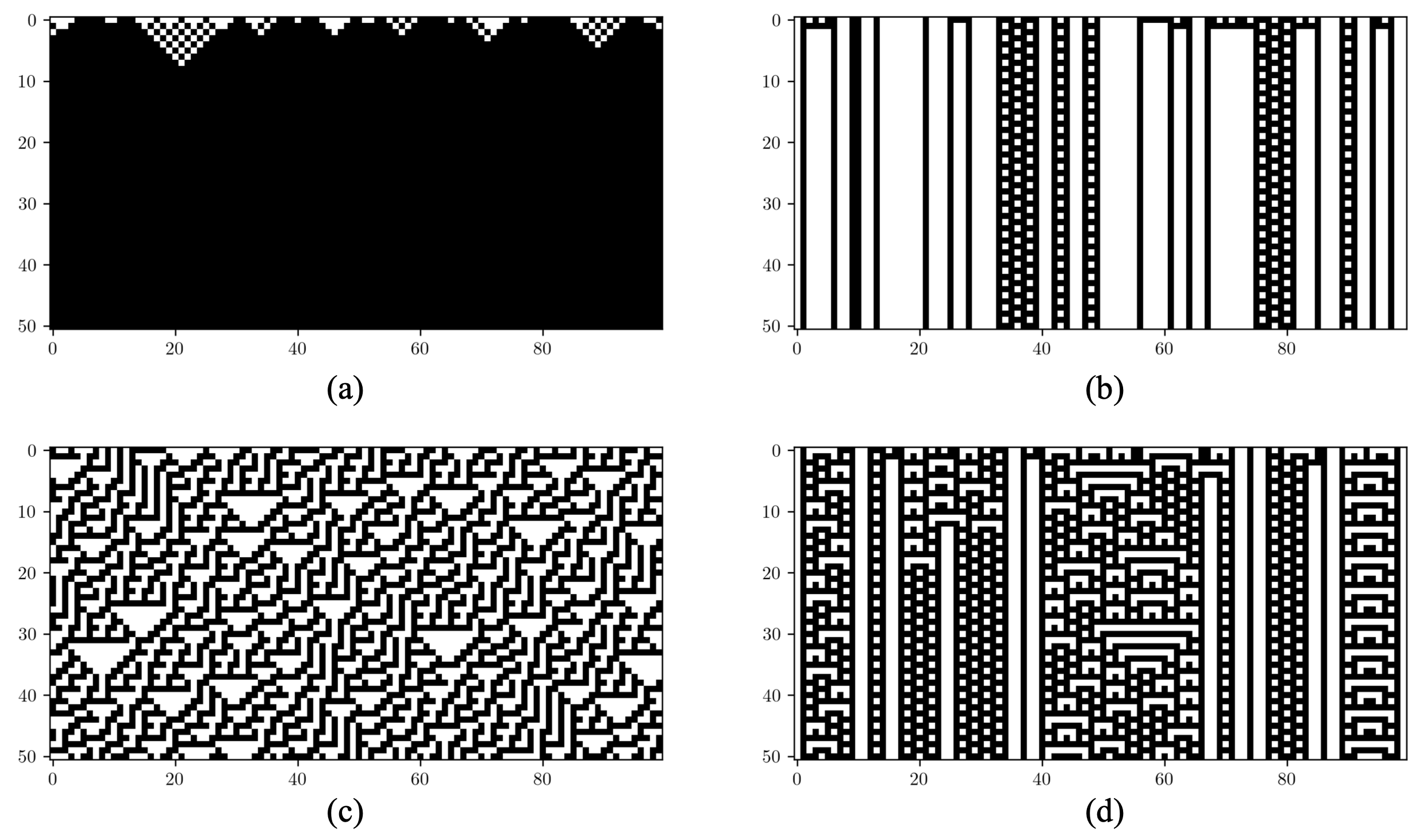} 
\caption{Examples of classes in elementary CAs. (a) Class 1. (b) Class 2. (c) Class 3. (d) Class 4. CA is a discrete formulation based on a regular uniform lattice or array with discrete values at each element. The actualization of each array depends on particular rules related to the neighborhood of each element. The y-axis displays the actualization of each array from top to bottom. The x-axis shows each element of the array.}
\label{classesCAs}
\end{figure}

Price coordination is a complex phenomenon that has been studied by different schools of economic thought. Even though economists have contributed to social sciences by understanding and describing many complex phenomenon of social interactions, there is no scientific consensus of what are the simplest mechanisms behind the price coordination. This problem is directly related to how we process and interpret the reality based on beliefs and objective information. For example, at one hand, we can generate theories that are internally consistent, but they are far from reality or incomplete. On the other hand, we can use empirical evidence to generate theories \citep{Barabasi2012}, but they can be internally inconsistent. Therefore, in the effort to explain complex phenomena of social coordination, each school of thought has suggested particular mechanisms or none at all. Nowadays, computational models can assist to approach a solution to this kind of open questions. In particular, the Wolfram's elementary cellular automata (CA) are computational models---one dimensional programs---for describing rich and complex behaviors based on simple rules that can be possible related to fundamental theories of social interactions \citep{vonNeumann1963, vonNeumann1966, vonNeumann1970,Wolfram2002, Wolfram2020}. There are fourth classes of elementary CAs based on a Wolfram's classification \citep{Wolfram2002} (Figure \ref{classesCAs}). 
Class 1 is related to uniform final states, and the class 2 shows different final states in which simple structures remain the same or repeat in particular time steps. Both of them show structures that display no further activity. Furthermore, class 3 is related to a marked presence of complex structures that can show small-scale patterns. In this class, we can see structures that change constantly, maintaining a high level of activity. Class 4 is a mix of class 2 and 3 in which there is a strong presence of random structures and localized structures. Consequently, the dynamics of these elementary CAs can replicate a wide range of natural and social behaviors because CAs are intended to be fundamental for explaining different phenomena. 

However, even though CAs are being highly used and analyzed in different disciples \citep{Turing1952, NicolisandPrigogine1977}, there is limited evidence of how to apply them in social sciences and humanities \citep{Hegselmann1996, LugoAlatristeContreras2022, LugoAlatristeContreras2024}. Therefore, in this brief study we aim to propose the use of elementary CAs related to economics. It often emphasizes a modeling approach over other methods used in social sciences. Even though
complex behaviors cannot always be completely capture in a system of equations with analytical solutions, computational models allow exploring all possible solutions. Such models are based on rules that describe micro level interactions that generate simulations showing trajectories and revealing dynamic consequences of the model. In this way, the economic theory based on the concept of rationality and its derivations has the potential for using CAs with the long tradition of formulating feasible types of coordinations between individuals and their collectivities that generate patterns for allocating different types of resources and satisfying needs and beliefs.

Generally speaking, studies of social sciences and humanities deal with the analysis of how the interaction of individuals or agents can generate a wide range of emergent behaviors that show observable patterns. Such patterns can be related to simple abstract ideas---e.g.,cultural dissemination \citep{Axelrod1984}---to complex real-life structures---e.g. a system of cities \citep{Krugman1996}. The key difference between both disciplines has been the methodological approach---i.e., quantitative vs qualitative methods. However, nowadays, the complexity approach offers a flexible framework for merging different ideas and methods of other disciplines based on a computational formulation. 
In particular, the computation based on programming languages provides the bases for exploring the formation of complex behaviors based on simple models. Therefore, in this study we are interested in 
using elementary CAs for describing the foundational principles of the economic theory based on the classical and neoclassical economics. We are interested in answering the following questions:
Which class of the elementary CA can be related to the dominant economic theory? and What are the rules in a elementary CA that describes different approaches of the economic theory? We hypothesize that the current classical and neoclassical approaches were build as a reductionist system searching for uniform or repetitive patterns of prices---stable or unstable equilibria---due to the lack of accesible computational power in that period of time. 
This type of limitation incorporated a set of assumptions and mathematical restrictions in which a feasible solution of problems was the only valid answer to complex phenomena. 
To quantify and validate such repetitive or inestable patterns based on restrictive assumptions, we aim to use an explorative data analysis based on the Shannon entropy and a conceptual model validation \citep{Shannon1948,LeekandPeng2015}. The former suggests the identification of uncertainty levels of different patterns of price coordination, and the latter identifies the significance of deterministic outputs of prices based on particular null models---price coordination based on pure guessing \citep{GotelliandGraves1996,Farine2017}.
Therefore, the economic theory can be translated as a set of simple rules that generate low levels of uncertainty, class 2, or high levels of uncertainty, class 4, of the elementary CAs.
The main contribution of this study is to rethink the economic theory by identifying the facts and believes behind the classical and neoclassical economics.

This document is dived in fourth sections. The first section is the literature review that describes, in broad terms, the fundamentals of the economic theory based on the classical and neoclassical approaches of economics. The second section shows the materials and method for building the elementary CAs and their simplest relationship with the economic theory. The third section displays the results. The last section discusses our findings and shows our conclusions.

\section{Literature review}
Contributions for developing economic theory can be traced from the different schools of economic thought \citep{Becchettietal2020}. In particular, we are interested in the classical and neoclassical economics because they exemplify the transition from early economics based on simple, logical, and heuristic formulations to a mathematical modeling of individual rationality and markets \citep{ColanderandKupers2016}. Behind these different approximations in the economic theory, there is a constant discussion about the type of participation of the government into the market. For example, on one hand, some economists support the idea that the free will of individuales---ultrarational persons---can control and guide the market. On the other hand, some economist are in favor of the idea that the government should provide all the principles---rights and rules---for coordinating the market. However, in a real context there is a mix of both in which self-organized and centralized decisions of agents can coordinate the market or other systems. Therefore, classical and neoclassical economics provide the bases for understanding key concepts for using an elementary CA.

The classical approach considers the rights of individuals for taking their own decisions based on moral beliefs as a key element for a successful social coordination. These ideas were related to the work of Adam Smith \citep{Smith1759,Smith1776}. He suggested that moral principles such as sympathy can coordinate actions of persons for generating a collective interest instead of looking at self-interests. Consequently, the concept of ``trust'' plays a key role for describing not only a social system, but also an economic system in which the common sense of a person related to personal and social conventions guides interactions. In addition, John Stuart \citet{Mill1848} went one step further using the concept of ``rationality'' as the basis for explaining the coordination of markers. He described the behavior of individuals based on objective decisions---verifiable facts, data, and evidence---related to a logical analysis. Even though this approach can be seen as an organized reasoning for being self-consistent, it shows important limitations in social and economic predictions. This approach describes the coordination of markets as a complex phenomena, i.e., the whole is not the sum of the parts. Therefore this classical vision stated that not only the economic, but also the society can be considered as a complex system in which simple principles related to subjective and objective decisions can guide a collective benefit. 

The neoclassical approach was the logical forward step in the classical vision. The mathematical formalization---logical concepts via symbols \citep{Frege1879, Siekmann2014}---became more precise in expressing complex relationships.
Economists such as Alfred \citet{Marshall1890}, and A.C. \citet{Pigou1920} used an organized reasoning based on objective principles for describing feasible solutions. This approximation considered the existence of analytic solutions based on a logical coherence in which normative and reasonable judgments---government and persons principles respectively---for coordinating the market were translated to a mathematical language. In particular, the method of analysis changed from a pure descriptive approximation, which considers the presence of analytic and non analytic solutions, to a formal mathematical analysis, which only considers analytic solutions. In this respect the work of \citet{Walras1896} and his followers---\citet{Pareto1909, Arrow1951,Debreu1951}---exemplify such a change based on a general equilibrium approach. In short, based on a large number of heroic implicit assumptions, the information of prices is the key element for knowing the presence of a market coordination. That is, prices of goods can change without any market logic, but at the end the Walras principle holds true---i.e., a constant balance between demand and supply in which the total demand equals total supply for all commodities.

Following this trend of applying the best available method and tools for contributing and developing the current scientific knowledge, we can use the computation of complex systems for exploring key ideas related to those schools of economic thought \citep{Kirman2010,WilsonandKirman2016}. In particular, using the current knowledge of programming language \citep{Wolfram2002, Wolfram2020}, we can identify the types of simple rules that generate patterns of coordination related to such schools of economic thought. In this respect, there is a coming back to the classical approach in the sense of exploring analytic and non analytic solutions based on the use of computation. 
Therefore, based on simple programming codes, we could model basic economic formulations and explore their current relevance and consistency in describing the coordination of markets and people.

The next parts of this study will describe elementary CAs that can explain the basic ideas of coordination based on the classic and neoclassic approaches.

\section{Materials}
The main source of materials for modeling elementary CAs is the work of \citet{Wolfram1983,Wolfram2002}. Based on these contributions, we generated our programming code that replicates one-dimensional CA. On this point, we used the Python programming language and their libraries for modeling, computing, and analyzing data. In particular, we used the following  third-party libraries: \href{https://matplotlib.org/}{https://matplotlib.org/} and \href{https://numpy.org/}{https://numpy.org/} \citep{Hunter2007,Harrisetal2020}. Because of the transparency, openness, and reproducibility of science \citep{Noseketal2015}, we share the code and data in the Open Science Framework (project: \href{https://osf.io/mgkqr/overview?view_only=94a7e7cde4234d249c93fd4a42115763}{Cellular automata and economic theory}).

\section{Method}
Following the main contributions of the economic theory related to the classical and neoclassical approaches, we propose a version of the elementary CA recommended by \citet{Wolfram2002} to understand and describe such schools of thought. In particular, the key point of understanding the coordination problem of prices in both of them, is related to identifying large-scale patterns of prices. For example, the classical approach suggested that the mechanism behind the coordination of prices is the rationality of the individual. Meanwhile, the neoclassical approach suggested that the individual takes the information of prices after the individual assumes the coordination of them. That is, the former suggests that prices are endogenously coordinated, and the latter proposed that prices are exogenously coordinated. Both have significant impact on understanding the coordination of individuals in different escenarios. 

The classical approach assumes a rational individual who is acting in self-interest for consuming or producing goods and generates a collective and positive coordination. This type of rationality proposed the use of objective data and logic information for making decisions. Therefore, this approach is interested in identifying what are the mechanisms behind such coordination between individuals and prices. For example, one of the key ideas of \citet{Marx1987} was to describe the price of goods based on their value, which is determined by the labor-time required to produce such goods. Beyond this approach, it is important to mention that this type of study is a current trend in the analysis of social sciences. For example, the work of \citet{Schelling1969, Schelling1978} demonstrated that people who are self-considered tolerant towards others can generate patterns of segregation. The work of \citet{Axelrod1984} explained that if a person shows an increase interaction between surrounding persons who share one of their beliefs, attitudes, and behavior, it is more likely that the person will adopt one of those traits. Consequently, there is a pattern towards a social influence convergence \citep{Axelrod1997}. The study of \citet{Krugman1996} proposed a ``racetrack economy'' that simulates the concentration of the economic activity into cities. Furthermore, the work of \citet{LugoMartinezMekler2022} suggested that Krugman's model can describe  the formation of a system of cities---their spacial location and size---if ports are spatially considered in the formulation. Therefore, the classical approach not only investigates the formation of stable patterns in which simple structures remain the same or repeat in particular discrete intervals, but also the presence of complex structures based on small-scale patterns or a strong presence of random structures and localized structures.

On the other hand, the neoclassical approach assumes a rational individual who is reacting by an hypothetical situation of a market balance, which is taken as existing. There is a lack of interest for describing and identifying mechanisms behind such a balance. In this respect, the work of \citet{Walras1896} proposed that prices were coordinated by their interactions across all markets moving the system toward a balance. That is, mechanisms for coordinating prices of goods are unknown. We can only assume that if we identify in a large-scale some stable or repetitive pattern in prices, we confirm the presence of a coordination. On the other hand, if we identify in a large-scale the presence of complex structures based on small-scale patterns or a strong presence of random structures and localized structures, we cannot confirm the presence of a coordination in the sense of the neoclassical approach. Therefore, this approach reduces the range of possible escenarios to those that are feasible.

Therefore, to translate the ideas of both approaches into an elementary CA, we propose to use the relationship between princes based on the concepts of  substitute and complementary goods \citep{MasColell1995}. In the following subsection, we describe the use of these concepts into modeling elementary CAs.

\subsection{Price interactions and elementary CAs}
Following the work of \citet{Wolfram2002, Wolfram2020} and \citet{Walras1896}, we can identify the rules of prices that correspond to the classical and neoclassical approaches in a one-dimensional cellular automata. Based on the \href{https://mathworld.wolfram.com/ElementaryCellularAutomaton.html}{Wolfram Mathworld} online documentation, we can set the value of each cell as (0 or 1), where 0 is a decreased price, and 1 is an increased price. Therefore, each cell represents the price of a good and its value characterizes the price adjustments based on its neighbors for balancing the market. 

\begin{figure}[ht]
\centering
\includegraphics[width=12.5 cm]{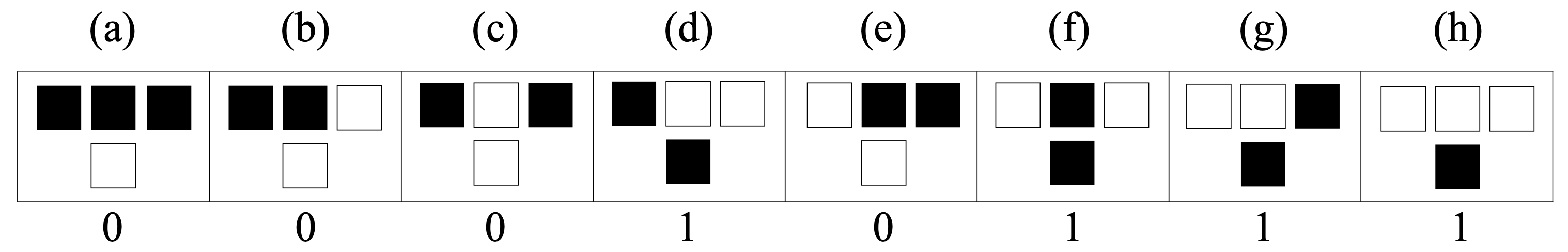} 
\caption{Price interactions and their actualization based on  a market balance.}
\label{figClassicalRules}
\end{figure}

The rules of a given cell and its actualization are based on the value of the cell to its right (substitute good) and its left (complementary good). Figure \ref{figClassicalRules} exemplifies the use of price interactions and their actualization. 

Based of the assumption that the movements of prices should balance the market, each combination of prices in Figure \ref{figClassicalRules} can be described as follows:
\begin{enumerate}
	\item [(a)] If a cell shows an increased price of a good and its right and left cells show the same behavior, the actualization of the cell for the next iteration will be a decreased price.
	\item [(b)] If a cell shows an increased price of a good and its left cell (complementary good) shows the same behavior, the actualization of the cell for the next iteration will be a decreased price.
	\item [(c)] If a cell shows a decreased price of a good and its right and left cells show a behavior of increased price, the actualization of the cell for the next iteration will be the same as the current state, i.e., a decreased price.
	\item [(d)] If a cell shows a decreased price of a good and its left cell (complementary good) shows an increased price and its right cell (substitute good) displays a decreased price, the actualization of the cell for the next iteration will be an increased price.
	\item [(e)] If a cell shows an increased price of a good and its right cell (substitute good) shows the same behavior, the actualization of the cell for the next iteration will be a decreased price.
	\item [(f)] If a cell shows an increased price of a good and its right and left cells show a behavior of decreased prices, the actualization of the cell for the next iteration will be the same as the current state, i.e., an increased price.
	\item [(g)] If a cell shows a decreased price of a good and its left cell (complementary good) shows a decreased price and its right cell (substitute good) displays an increased prince, the actualization of the cell for the next iteration will be an increased price.
	\item [(h)] If a cell shows a decreased price of a good and its right and left cells show the same behavior, the actualization of the cell for the next iteration will be an increased price.
\end{enumerate}
 
Therefore, based on the classical or neoclassical approach, we can set the type of rule behind the main theoretical foundations of price coordinations.
 
\subsection{Shannon entropy}
This theoretical study is based on an explorative data analysis that aims to show new ways of use of elementary CAs and the fundamentals of the economic theory. In this respect, one helpful mesure for computing the order or disorder in systems is the Shannon entropy \citep{Shannon1948}. In our context, we are interested in measuring the uncertainty related to the classical and neoclassical approaches by using the well-known elementary CAs. We compute the Shannon entropy following the next formulation:

\begin{equation}
\label{eq:2}
	H = -\sum_{i=1}^{n} p_{i} log_{2} p_{i}
\end{equation}
in which $H$ is the entropy, $p_{i}$ is the probabilities of black and white cell $i$ associated with the \emph{n}-items of given intervals $n$. Therefore, decreasing values of $H$ indicate high uncertainty. 

In this respect, it is important to highlight that we used the term of ``uncertainty'' instead of ``diversity'' associated with the entropy. The former is related to physical and the latter to biological sciences \citep{Jost2006}. In this study, we are interested in describing the entropy based on the frequencies of the data instead of the number of bins that cover the range of such a data. Therefore, the Shannon entropy can provide a measurement for identifying how the theoretical bases of the classical or neoclassical economics can generate particular levels of uncertainty associated with the price coordination. For example, equilibrium patterns can be related to low levels of uncertainty, meanwhile disequilibrium or multiple equilibrium patterns can be related to high levels of uncertainty.

\subsection{Model validation}
We use a conceptual model validation in which particular rules of elementary CAs are compared with singular null models for identifying whether such deterministic rules deviate from random chance \citep{KerrandGoethel2014}. 
Based on a Monte Carlo simulation approximation, we use three types of null models to determine if our results could have arisen by random chance. 
According to the work of \citet{MolugaramandRao2017} and \citet{Lugoetal2025}, we use a binomial distribution with different values of the parameter $p$---probability of success---as a baseline model for comparison. This distribution is related to the number of successes in a fixed number of independent Bernoulli trials. The Bernoulli distribution is a discrete probability distribution in which a random variable $X$ can take only two values $X = 1$ for success and $X = 0$ for failure. In particular, we are interested in a sensitive analysis of the probability of success when p = 0.3, p = 0.5, and p = 0.7. The first case describes a 30\% chance for selecting an increase in price. The second and third cases define 50\% and 70\% chances for selecting an increase in price. Consequently, each selected rule of the elementary CAs shall be compared with the median of $10,000$ realizations per case by computing the Spearman correlation coefficient. This type of comparison using the Spearman correlation can identify how statistically significant or statistically different from random chance is each deterministic rule. 
Therefore, these cases explore randomized escenarios in which deterministic rules can be related to biased or unbiased selection of increase or decrease prices. 

\subsection{Initial conditions}
We used an array in one-dimension with a length of $100$. This array was generated based on a wraparound boundary conditions or torus. Each item of this array is associated with value of 0 or 1. Based on the work of  \citet{Wolfram2002} we used two types of visualizations for showing the dynamics: a) random initial conditions in which values of 0 and 1 are selected based on a uniform distribution, and b) initial state with a 1 in the middle of the array. We display the dynamics of 50 actualizations, from top to bottom, in which each updated (row) is related to the previous array. It is important to mention that after this number of iterations, the model shows patterns that can be clearly identified by one of the fourth classes of elementary CAs, compute the Shannon entropy, and validate the model.

\section{Results}
The aim of this study was to identify the type of elementary CA related to the classical and neoclassical approaches of the economic theory. The process of evaluating and showing this identification is not trivial. We used a theoretical approximation in which fundamentals of the classical and neoclassical economics were applied into simple rules of elementary CAs. Bellow we show our main results. 

Based on the classical approach of the economic theory, we could identify three rules of prices that are consistent with a rational choice based on a deductive reasoning and complete information. The first result is related to the rule $000101111_2$ in the classification of \citet{Wolfram2002}. 

\begin{figure}[!h]
\centering
\includegraphics[width=10.5 cm]{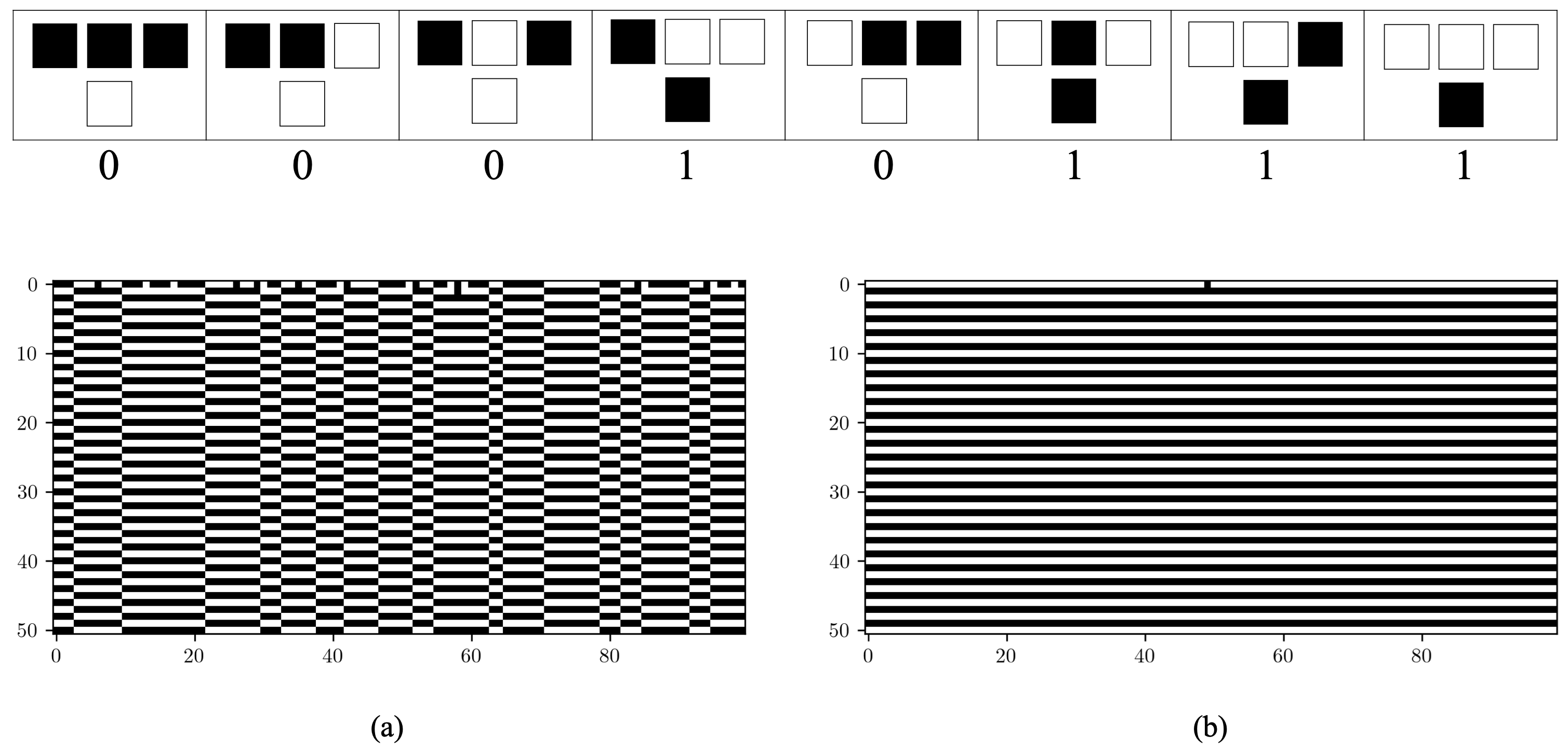} 
\caption{Classical Rule 01. (a) random initial condition. (b) simple initial condition.}
\label{res01Classical}
\end{figure}   
Figure \ref{res01Classical} shows stable patterns in (a) and (b). Prices are balanced based on a rational logic in which the presence of three or two similar price behaviors at a current state produce an oposite behavior in the next iteration. In terms of \citet{Wolfram2002} classification, Figure \ref{res01Classical} shows the rule $23 = 00010111_2$. 
Table \ref{nullmodels01} shows that (a) and (b) cases are statistically different from a random binomial variable, meanwhile the entropy values confirm highly structured patterns. For simplicity and easy of reading the following Tables, we display a reduce version of Table \ref{nullmodels01} caption.  
Therefore, this rule gives the same importance of prices of substitute and complementary goods. 

\begin{table}[!h]
\centering
\begin{tabular}{|c|c|c|c|c|}
\hline
{\bf Classical rule 01} & {\bf p=0.3} &{\bf p=0.5} & {\bf p=0.7} & {\bf Entropy}\\
\hline
(a)	& (-0.1386, 0.3318)	& (-0.1386, 0.3318)	& (-0.1386, 0.3318) & 0.9975 \\
(b)	& (0.0, 1.0)	& (0.0, 1.0)	& (0.0, 1.0) & 0.9999\\
\hline
\end{tabular}
\caption{\label{nullmodels01}{\bf Model validation and entropy in the Classical rule 1}. The model validation is related to the correlation between escenarios and null models. Columns with $p=0.3$, $p=0.5$, and $p=0.7$ show the probability of success for selecting 1s. Values in tuples show the Spearman correlation between a deterministic rule and our proposed null models based on the binomial distribution in a fixed number of independent Bernoulli trials. Each tuple shows the Spearman correlation coefficient and its p-value. This measure computes the value associated with each deterministic escenario with the median of $10,000$ realizations per null model of the number of 0s. The probability density of the binomial distribution is the following: $p(N)=\tbinom{n}{N} p^{N}(1-p)^{n-N}$ where $n=$number of trails, $p=$probability of success, and $N=$number of successes. The probability mass function of the Bernoulli distribution is the following: $f(k)=\{1-p \quad if \quad k=0, p \quad if \quad k=1 \}$ for k in $\{0,1\}$, $0<= p <= 1$.
}
\end{table}

\begin{figure}[!h]
\centering
\includegraphics[width=10.5 cm]{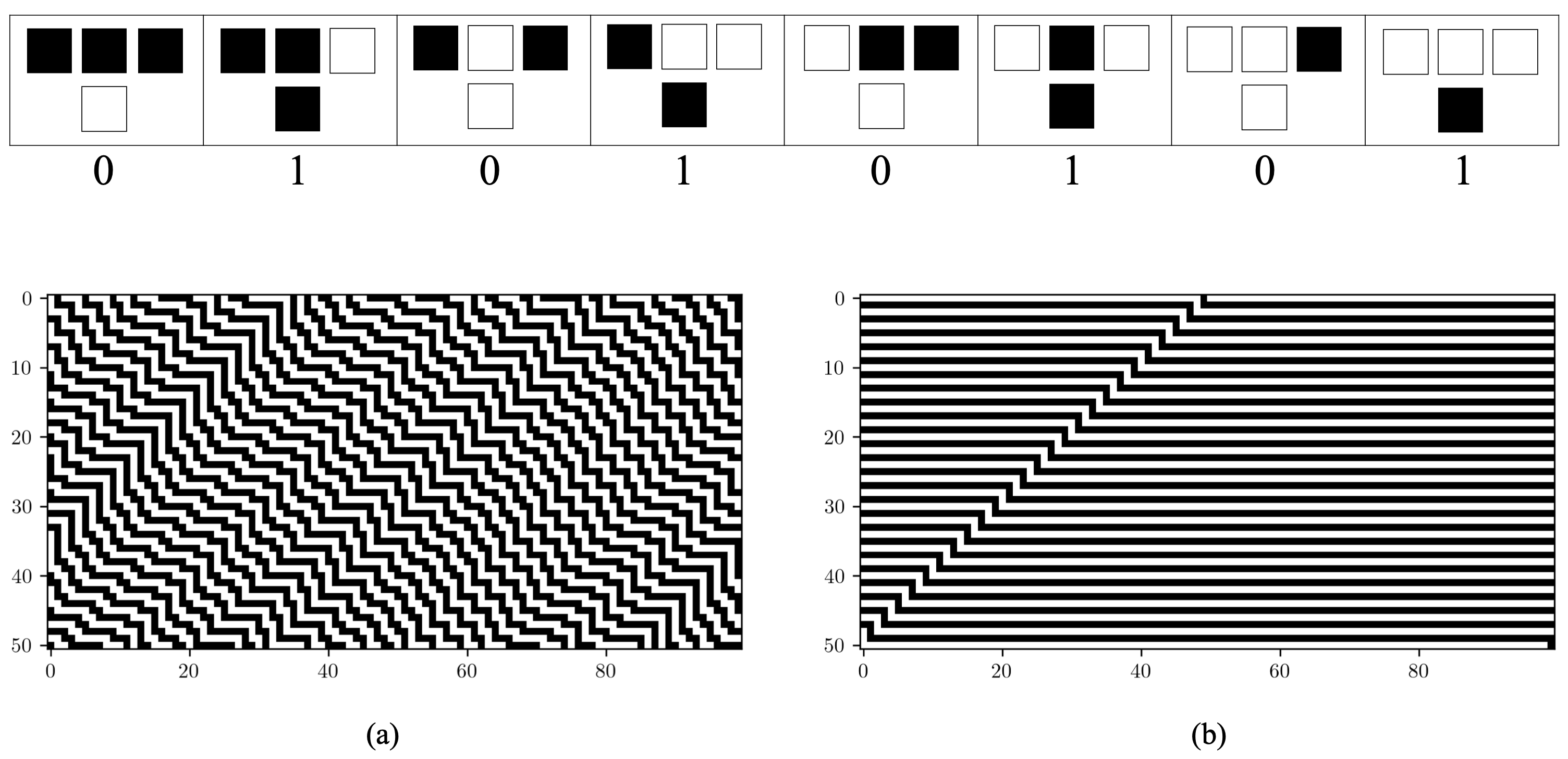} 
\caption{Classical Rule 02. (a) random initial condition. (b) simple initial condition.}
\label{res02Classical}
\end{figure}

Next, Figure \ref{res02Classical} displays a variation related to the price of substitute goods. This variation intends to break the symmetry between complementary and substitute goods by giving more importance to the presence of the substitute goods.

Figure \ref{res02Classical} shows stable patterns in (a) and (b). Prices are balanced based on a rational logic in which the presence of a substitute determines the actualization in the next iteration. In terms of \citet{Wolfram2002} classification, this is the rule $85 = 01010101_2$ that gives the price of a substitute more importance than complementary goods. In addition, 
Table \ref{nullmodels02} shows that (a) and (b) cases are statistically different from a random binomial variable, meanwhile the entropy values confirm highly structured patterns. 

\begin{table}[!h]
\centering
\begin{tabular}{|c|c|c|c|c|}
\hline
{\bf Classical rule 02} & {\bf p=0.3} &{\bf p=0.5} & {\bf p=0.7} & {\bf Entropy}\\
\hline
(a)	& (0.1386, 0.3318)	& (-0.1386, 0.3318)	& (-0.1386, 0.3318) & 0.9999 \\
(b)	& (-0.1386, 0.3318)	& (0.1386, 0.3318)	& (0.1386, 0.3318) & 0.9997\\
\hline
\end{tabular}
\caption{\label{nullmodels02}{\bf Model validation and entropy in the Classical rule 02}. 
}
\end{table}

\begin{figure}[!h]
\centering
\includegraphics[width=10.5 cm]{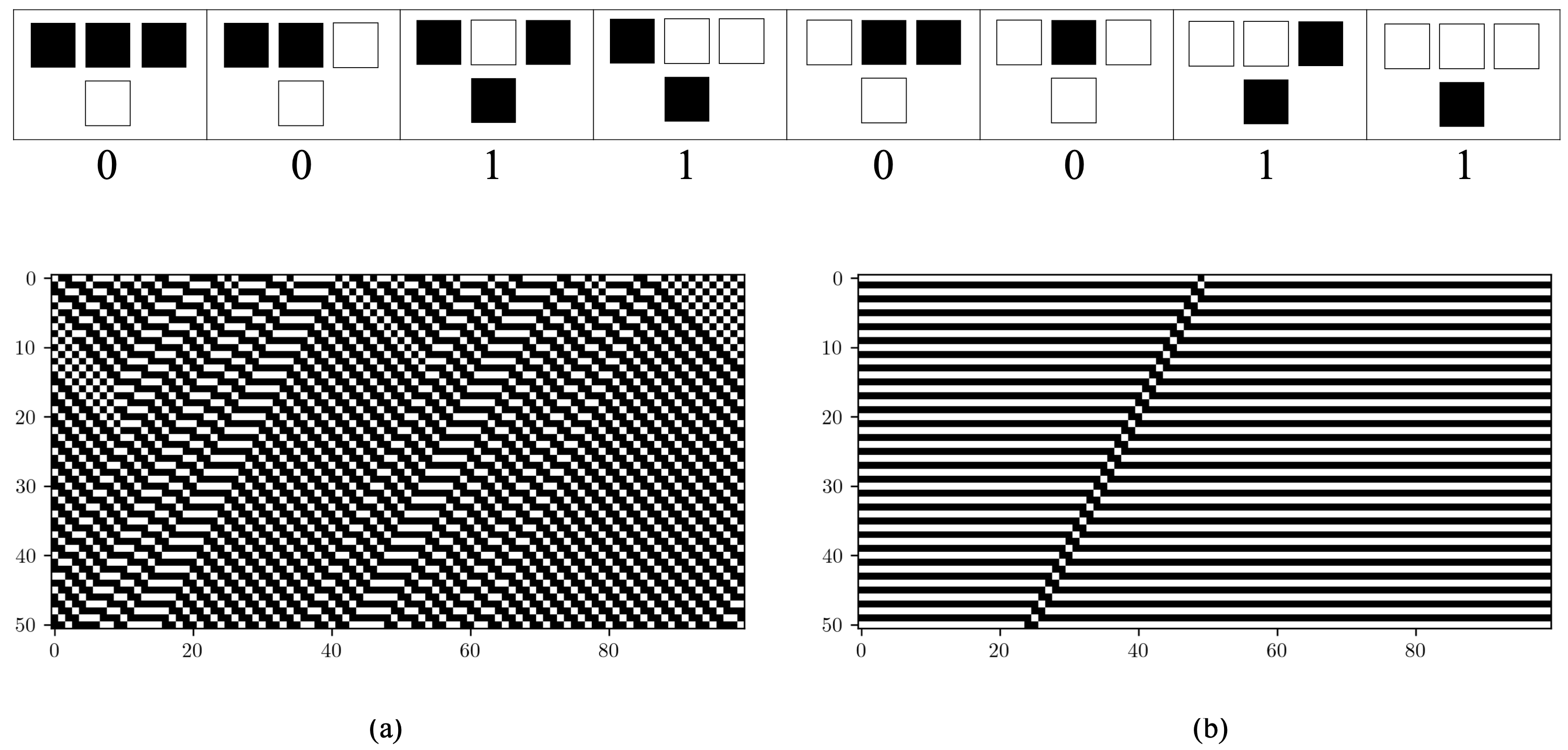} 
\caption{Classical Rule 03. (a) random initial condition. (b) simple initial condition.}
\label{res03Classical}
\end{figure}

Subsequently, Figure \ref{res03Classical} displays a variation related to the price of a complementary goods. This variation describes a situation in which, if the price of one good increases, the demand of its complements falls.

Figure \ref{res03Classical} shows a complex pattern in (a), but (b) shows a stable pattern. Prices are balanced based on a rational logic in which the presence of a price of a complementary good determines the actualization in the next iteration. In terms of \citet{Wolfram2002} classification, this is the rule $51 = 00110011_2$ that gives the price of a complementary good more importance than the price of substitute goods. In addition, 
Table \ref{nullmodels03} shows that (a) and (b) cases are statistically different from a random binomial variable, meanwhile the entropy values confirm highly structured patterns. 

\begin{table}[!h]
\centering
\begin{tabular}{|c|c|c|c|c|}
\hline
{\bf Classical rule 03} & {\bf p=0.3} &{\bf p=0.5} & {\bf p=0.7} & {\bf Entropy}\\
\hline
(a)	& (0.1386, 0.3318)	& (-0.1386, 0.3318)	& (-0.1386, 0.3318) & 0.9628 \\
(b)	& (-0.1386, 0.3318)	& (0.1386, 0.3318)	& (0.1386, 0.3318) & 0.9999\\
\hline
\end{tabular}
\caption{\label{nullmodels03}{\bf Model validation and entropy in the Classical rule 03}. 
}
\end{table}

\begin{figure}[!h]
\centering
\includegraphics[width=10.5 cm]{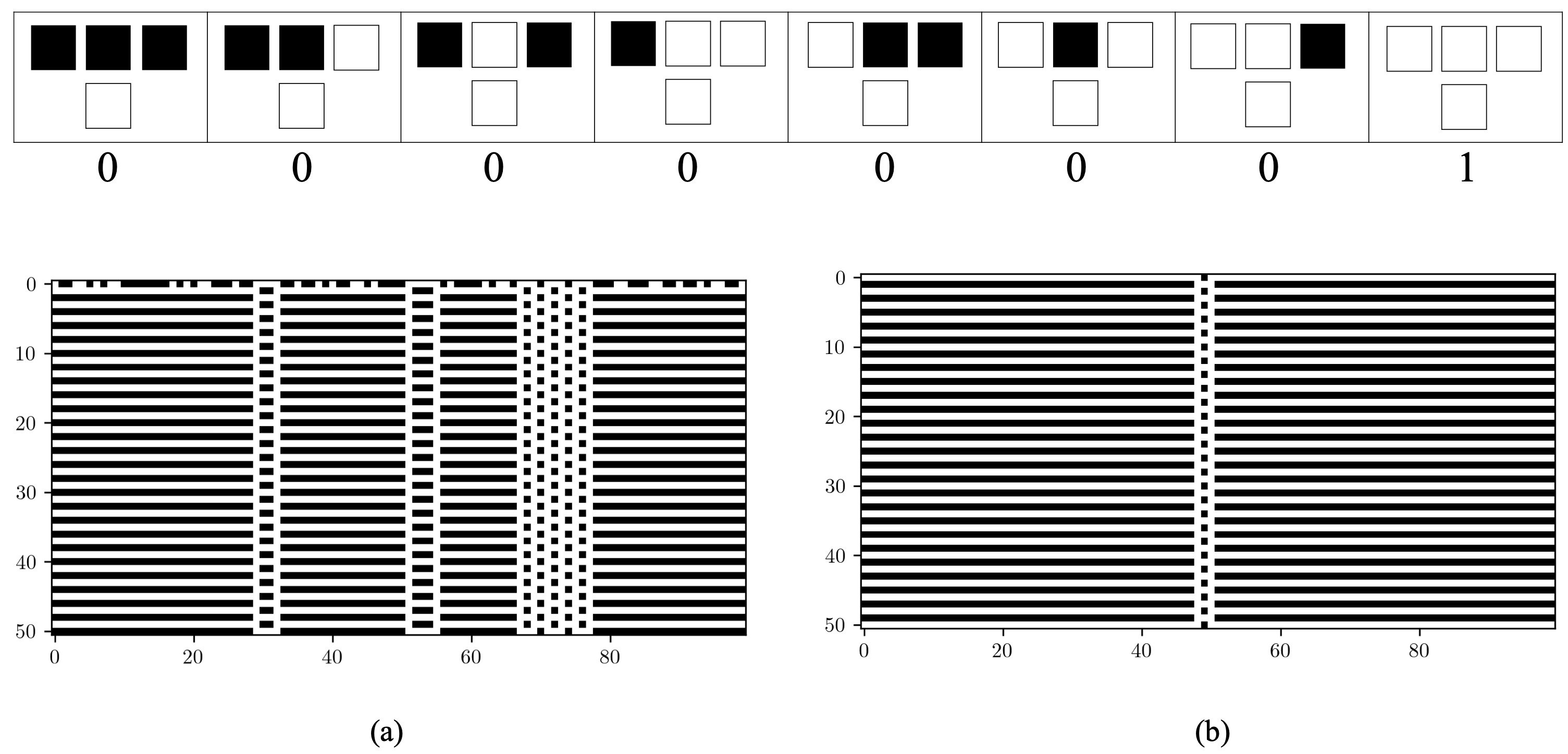} 
\caption{Examples of class 1 of elementary CAs. Rule 1. Table \ref{nullmodels001} shows that (a) and (b) cases are statistically different from a random binomial variable, and the entropy values confirm highly structured patterns.}
\label{rule001}
\end{figure}

\begin{figure}[!h]
\centering
\includegraphics[width=10.5 cm]{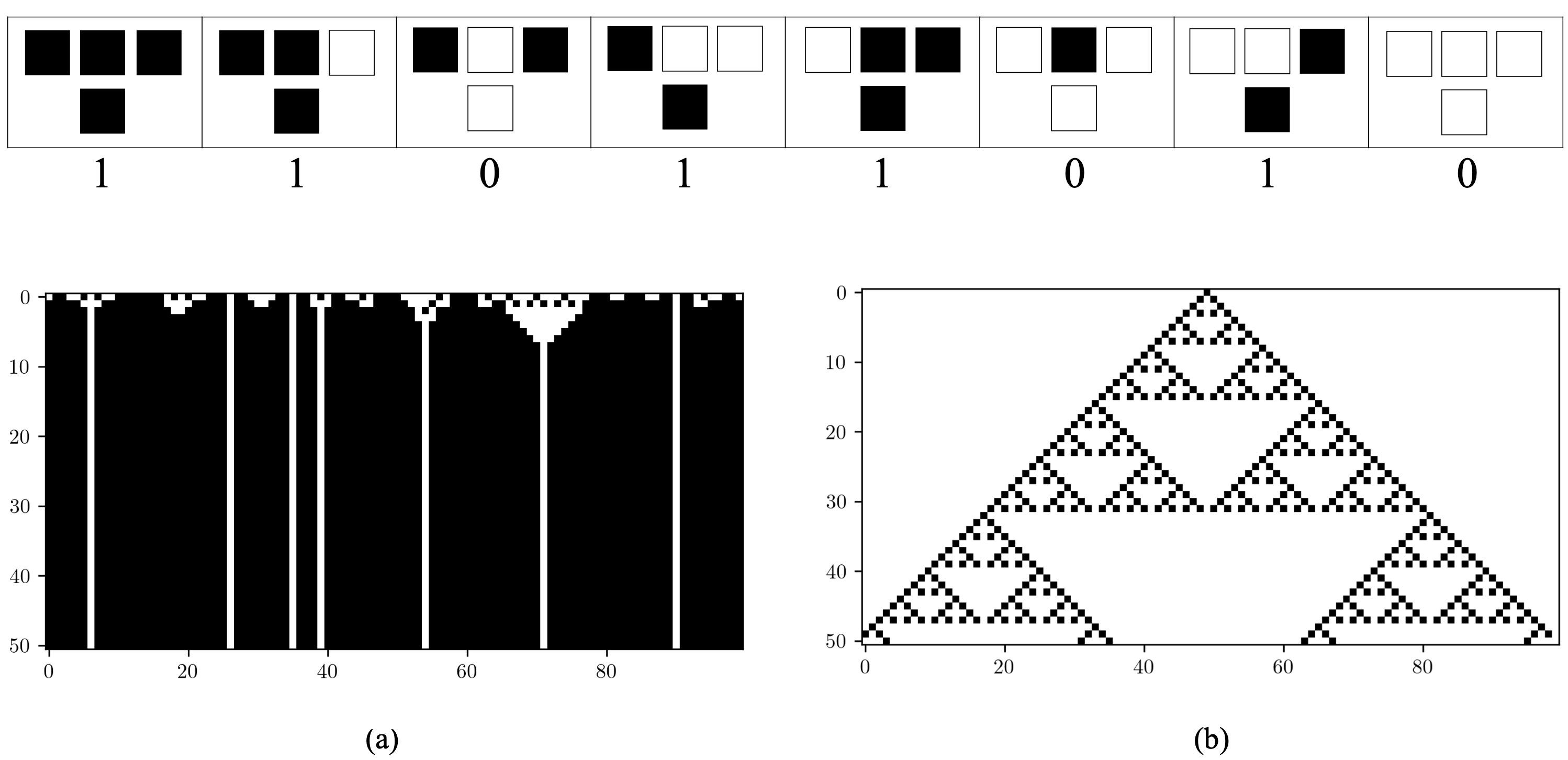} 
\caption{Examples of class 2 of elementary CAs. Rule 218. Table \ref{nullmodels218} shows that (a) and (b) cases are statistically significant at $p < 0.10$, and the entropy values confirm periodic interesting patterns over time different from a highly ordered system.}
\label{rule218}
\end{figure}

Turning now to the neoclassical approach, it can be exemplified by a total of $2^8 = 256$ rules described by \citet{Wolfram1983,Wolfram2002}. Because  the main assumption of this approach is that the coordination of prices is unknown, and an imaginary "auctioneer" announces prices when they are in equilibrium, we are only able to identify the type of large-scale patterns of price coordinations. Therefore, this approach is interesting only for identifying stable or repetitive large-scale patterns. In other words, the neoclassical approach is inclined to see price behavior related to class 1 and 2 of CAs (Figures \ref{rule001} and \ref{rule218}).   

\begin{table}[!h]
\centering
\begin{tabular}{|c|c|c|c|c|}
\hline
{\bf Rule 1} & {\bf p=0.3} &{\bf p=0.5} & {\bf p=0.7} & {\bf Entropy}\\
\hline
(a)	& (0.0, 1.0)	& (0.0, 1.0)	& (0.0, 1.0) & 0.9931 \\
(b)	& (-0.1386, 0.3318)	& (0.1386, 0.3318)	& (0.1386, 0.3318) & 0.9989\\
\hline
\end{tabular}
\caption{\label{nullmodels001}{\bf Model validation and entropy in the Rule 1}. 
}
\end{table}

\begin{table}[!h]
\centering
\begin{tabular}{|c|c|c|c|c|}
\hline
{\bf Rule 108} & {\bf p=0.3} &{\bf p=0.5} & {\bf p=0.7} & {\bf Entropy}\\
\hline
(a)	& (-0.4014, 0.0035)	& (0.4014, 0.0035)	& (0.4014, 0.0035) & 0.4358\\
(b)	& (-0.2489, 0.0780)	& (0.2489, 0.0780)	& (0.2489, 0.0780) & 0.4124\\
\hline
\end{tabular}
\caption{\label{nullmodels218}{\bf Model validation and entropy in the Rule 218}. 
}
\end{table}

\begin{figure}[!h]
\centering
\includegraphics[width=10.5 cm]{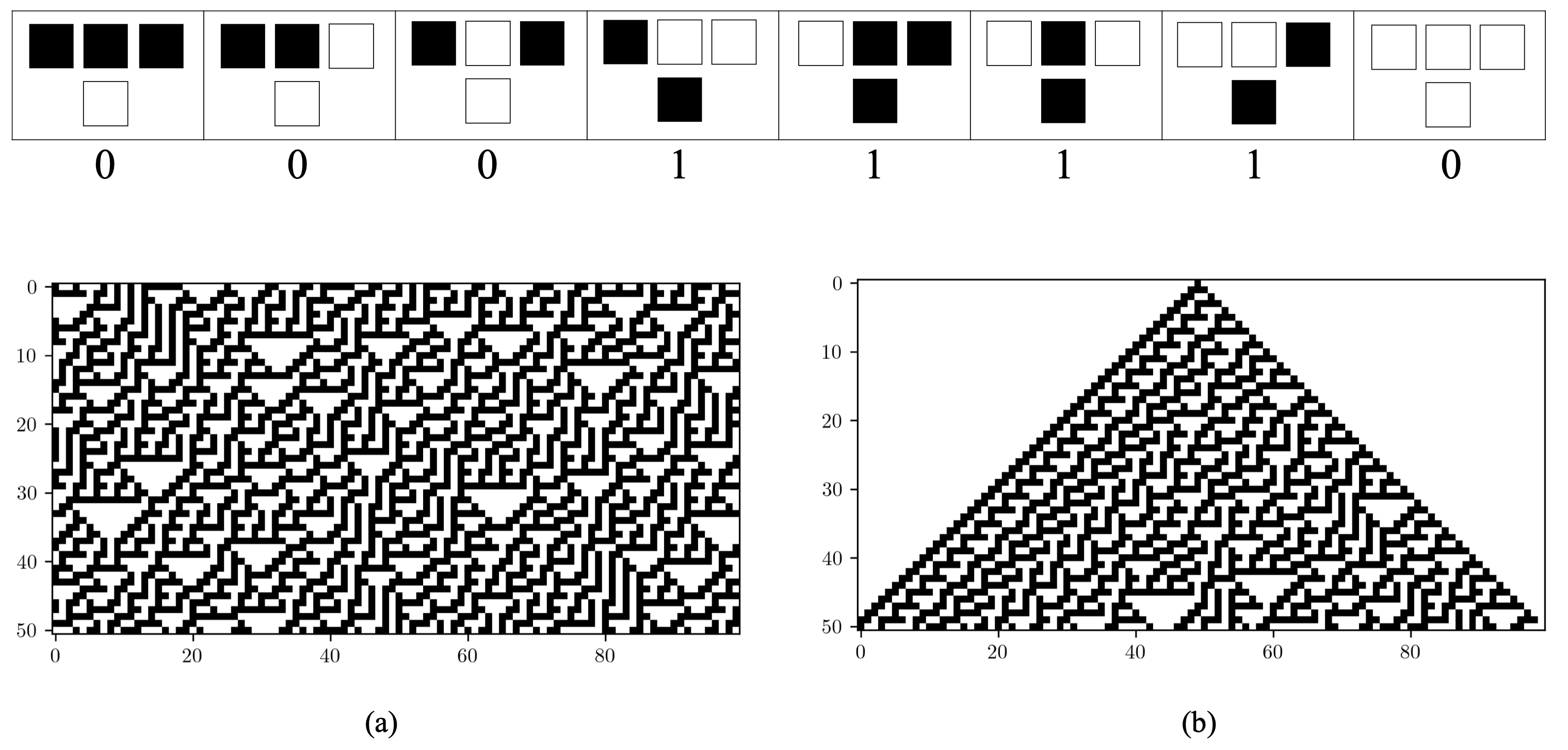} 
\caption{Examples of class 3 of elementary CAs. Rule 30. Table \ref{nullmodels30} shows that (a) is statistically different from a random binomial variable, but (b) is statistically significant at $p < 0.10$. The entropy values shows that (a) and (b) display pseudo-random patterns over time.}
\label{rule30}
\end{figure}

Based on the work of \citet{deSouzaCarvalho2009}, we can identify at least 150 rules related to stable or repetitive large-scale patterns. There is a $58.5\%$ of rules that generate possible patterns associated with price coordination in neoclassical economics. 
Consequently, the neoclassic economics is not interested in the classes 3 and 4 because it avoids non-equilibrium prices (Figure \ref{rule30}). 

\begin{table}[!h]
\centering
\begin{tabular}{|c|c|c|c|c|}
\hline
{\bf Rule 30} & {\bf p=0.3} &{\bf p=0.5} & {\bf p=0.7} & {\bf Entropy}\\
\hline
(a)	& (0.0818, 0.5677)	& (-0.0818, 0.5677)	& (-0.0818, 0.5677) & 0.9999\\
(b)	& (-0.2402, 0.0894)	& (0.2402, 0.0894)	& (0.2402, 0.0894) & 0.8414\\
\hline
\end{tabular}
\caption{\label{nullmodels30}{\bf Model validation and entropy in the Rule 30}. 
}
\end{table}

The presence of complex structures of prices or random behaviors that represents unstable patterns are analytically intractable. In this case, the neoclassical approach is unable to give any guide for identifying equilibriums, and even worse, to explore some mechanisms for price coordinations.

\section{Discussion}
Nowadays, it is fundamental to use the computation for clarifying current scientific ideas and developing original thoughts based on theoretical models and data evidence. Theoretical models based on computation are excellent tools for exploring logical escenarios that can explain the phenomena in question, as well as for describing alternative escenarios that can show logical and non-logical behaviors. In this case, the use of elementary CAs can help to describe different economic behaviors that are parte of the core of the economic theory. Therefore, these elementary CAs can describe some mechanism for coordinating the price.

The classical approach provides the bases for exploring the price coordination based on the intuition of social interactions. 
For example, if there is a cultural context in which is preferred to produce and consume some particular goods instead of other goods due to a social opinion based on rational beliefs, attitudes, and behaviors, there is a social influence that determines the success or failure of goods. In terms of the market, the fluctuation of prices is related to the dynamics of such a social opinion.

In the neoclassical approach, playing dice and coordinating prices is the same type of unpredictable event. The interaction between prices assumes that price interactions do not follow any type of direct social influence. Coordination of prices can emerge even if there is a complete random process behind price interactions. In particular, because the individual takes such information and reacts, this behavior is similar to when individuals react to the weather of a day. Individuals cannot influence directly the weather of a day, they just identify and react to some clear pattern. However, as we know, thanks to the climate change studies, individuals affect the weather through daily choices---i.e., emitting greenhouse gases. Therefore, under the neoclassical assumption of time preferences---individuals value immediate satisfaction more than delayed rewards---the indirect effect of individuals to prices is not theoretically relevant due to the time horizon. 
 
Based on the above discussion, we partially confirm our hypothesis that the economic theory is related to the class 2 and 4. We confirmed that the classical approach is more interested in price coordination related to classes 3 and 4. However, because of the lack of accesible computational power in that period of time, economists could only find logic solutions of price coordination when they identify pattern behaviors related to class 1 and 2. On the other hand, the neoclassical approach is interested exclusively in class 1 and 2. Therefore, a positive result of this is that both approaches complement each other in the economic theory. Even though the classical and neoclassical economics are different in their scientific formulation, economists should consider them for identifying accurately the type of method and tools to study any type of coordination problems.

One particular suggestion for students of economics is to recognize the facts versus believes behind the economic theory. In particular, economics has tried to describe logically and objectively complex phenomena related to systems generated by the human being. However, for simplicity in the analysis, economists have considered two types of approaches that have not fully identified simple mechanisms for describing certain types of coordinations. In particular, the price coordination is one of the ideas into the economic theory that can only exist under certain conditions but not in real life. Therefore, modeling the economic theory based on computational concepts, we could reveal such facts and believes. According to \citet{Kirman2010}, it is the coordination mechanism that is central in the economy. A first step to model is to recognize that the economy is a complex system.

A limitation of this study is the interpretation of the 256 rules in a social context. We could only identify the direct theoretical fundamentals of the classical and neoclassical economics. However, subsequent studies should investigate each elementary CA rules and their relationship with different social theories of coordination and variations of the economic theory. 

\subsection{Conclusions}
Elementary CAs are extraordinary tools for exploring the consistency of the different approaches of economic theory. They provide the options for modeling simple rules that generate simple or complex scenarios related to particular schools of economic thought. Elementary CAs not only assist for identifying and clarifying facts versus believes into the price coordination in the economic theory, but also the presence of coordination patterns in social sciences and humanities.

\section*{Disclosure statement}
The authors report there are no competing interests to declare.

\section*{Funding}
No funding was received for conducting this study.

\section*{Ethics approval}
This study did not require the local institutional review board (IRB) approval because this study does not include human research participants.

\section*{Consent to participate}
Informed consent for human research participants to be published in this study was not obtained because this study does not involve human participants.

\section*{Data availability}
The data is exclusive for scientific purposes. The data and the code that support the findings of this study are openly available in ``Cellular automata and economic theory'' at \href{https://osf.io/mgkqr/overview?view_only=94a7e7cde4234d249c93fd4a42115763}{the Open Science Framework (OSF)}





\begin{thebibliography}{99}

\bibitem[Barabási(2012)]{Barabasi2012}
Barabási, AL. (2012), The network takeover. Nature Phys 8, 14--16. \href{https://doi.org/10.1038/nphys2188}{https://doi.org/10.1038/nphys2188}

\bibitem[von Neumann(1963)]{vonNeumann1963}
von Neumann, J. (1963), The General and Logical Theory of Automata, in J.von Neumann, "Collected Works" (ed. A.H.Taub), Vol. 5, p.288--328, \href{https://www.vordenker.de/ggphilosophy/jvn_the-general-and-logical-theory-of-automata.pdf}{https://www.vordenker.de/ggphilosophy/jvn\_the-general-and-logical-theory-of-automata.pdf}

\bibitem[von Neumann(1966)]{vonNeumann1966}
von Neumann, J. (1966), Theory of Self-Reproducing Automata, (ed. A.W.Burks), Univ. of Illinois press (1966). \href{https://cba.mit.edu/events/03.11.ASE/docs/VonNeumann.pdf}{https://cba.mit.edu/events/03.11.ASE/docs/VonNeumann.pdf}

\bibitem[von Neumann(1970)]{vonNeumann1970}
von Neumann, J. (1970), Essays on Cellular Automata, ed. A.W.Burks, Univ. of Illinois press, 


\bibitem[Wolfram(2002)]{Wolfram2002}
Wolfram, S. (2002), A New Kind of Science, Wolfram Media, Inc.

\bibitem[Wolfram(2020)]{Wolfram2020}
Wolfram, S. (2020). A Class of Models with the Potential to Represent Fundamental Physics, Complex Systems, 29(2), 2020 pp. 107--536.
\href{https://doi.org/10.25088/ComplexSystems.29.2.107}{https://doi.org/10.25088/ComplexSystems.29.2.107}

\bibitem[Turing(1952)]{Turing1952}
Turing, A. M. (1952), The chemical basis of morphogenesis, Philos. Trans. R. Soc. London, Ser. B 237, 37.

\bibitem[Nicolis and Prigogine(1977)]{NicolisandPrigogine1977}
Nicolis, G. and Prigogine, I., (1977), Self-Organization in Nonequilibrium Systems (Wiley, New York).

\bibitem[Hegselmann(1996)]{Hegselmann1996}
Hegselmann, R. (1996). Cellular Automata in the Social Sciences. In: Hegselmann, R., Mueller, U., Troitzsch, K.G. (eds) Modelling and Simulation in the Social Sciences from the Philosophy of Science Point of View. Theory and Decision Library, vol 23. Springer, Dordrecht. \href{https://doi.org/10.1007/978-94-015-8686-3_12}{https://doi.org/10.1007/978-94-015-8686-3\_12}

\bibitem[Lugo and Alatriste-Contreras(2022)]{LugoAlatristeContreras2022}
Lugo, I., Alatriste-Contreras, M.G. (2022). Intervention strategies with 2D cellular automata for testing SARS-CoV-2 and reopening the economy. Sci Rep 12, 13481. \href{https://doi.org/10.1038/s41598-022-17665-3}{https://doi.org/10.1038/s41598-022-17665-3}

\bibitem[Lugo and Alatriste-Contreras(2024)]{LugoAlatristeContreras2024}
Lugo, I., Alatriste-Contreras, M.G. (2022). The personal space and the collective behavior of crowd disasters, BioRxive: \href{https://www.biorxiv.org/content/10.1101/2024.09.05.611443v2}{https://doi.org/10.1101/2024.09.05.611443}


\bibitem[Axelrod(1984)]{Axelrod1984}
Axelrod, R. (1984), The Evolution of Cooperation. New York: Basic Books.

\bibitem[Krugman(1996)]{Krugman1996}
Krugman, P. (2022). The Self Organizing Economy. Wiley-Blackwell.

\bibitem[Shannon(1948)]{Shannon1948} 
Shannon, C. (1948). A Mathematical Theory of Communication. Bell System Technical Journal 27(3).

\bibitem[Leek and Peng(2015)] {LeekandPeng2015}
Leek, J.T. and Peng, R.D. (2015),What is the question?\emph{Science}, 347,1314--1315(2015). \href{https://www.science.org/doi/10.1126/science.aaa6146}{doi:10.1126/science.aaa6146}

\bibitem[Gotelli and Graves(1996)]{GotelliandGraves1996}
Gotelli, N.J. and Graves, G.R. (1996) Null Models in Ecology. Smithsonian Institution Press, Washington, DC, USA.

\bibitem[Farine(2017)]{Farine2017}
Farine, D.R. (2017), A guide to null models for animal social network analysis, Methods in Ecology and Evolution 8, 1309--1320, \href{https://doi.org/10.1111/2041-210X.12772}{https://doi.org/10.1111/2041-210X.12772}



\bibitem[Becchetti et. al.,(2020)]{Becchettietal2020}
Becchetti, L., Bruni, L. and Zamagni, S. (2020).
Chapter 1 - Economics: What it studies, with what methods, and how it evolved, The Microeconomics of Wellbeing and Sustainability,
Academic Press, pages 1--49.


\bibitem[Colander and Kupers(2016)]{ColanderandKupers2016}
Colander, D, and Kupers, R. (2016), Complexity and the Art of Public Policy: Solving Society's Problems from the Bottom Up, Princeton University Press

\bibitem[Smith(1759)]{Smith1759}
Smith, A.(1756), The Theory of Moral Sentiments, London: A. Miller 

\bibitem[Smith(1776)]{Smith1776}
Smith, A.(1776), The Wealth of Nations, London: A. Miller 

\bibitem[Mill(1848)]{Mill1848}
Mill, J.S. (1848), Principles of Political Economy, London: Longmans, Green.

\bibitem[Marshall(1890)]{Marshall1890}
Marshall, A. (1890), Principles of Economics. 9th edition. London Macmillan.

\bibitem[Pigou(1920)]{Pigou1920}
Pigou, A.C. (1920), The Economics of Welfare London, Reprint. New Brunswick, N.J.: Transaction.

\bibitem[Walras(1896)]{Walras1896}
Walras, L (1896), Éléments d'économie politique pure; ou, Théorie de la richesse sociale, Lausanne, F. Rouge Editeur, Librairie de l'Universite  \url{http://digamo.free.fr/walras96.pdf}

\bibitem[Pareto(1909)]{Pareto1909}
Pareto, V. (1909), Manual of Political Economy: A Critical and Variorum Edition, Oxford University Press.

\bibitem[Arrow(1951)]{Arrow1951}
Arrow, K.J. (1951) An extension of the basic theorems of classical welfare economics, Proceedings of the Second Berkeley Symposium on Mathemalical Statisties and Probability, ed J. Neyman, University of California Press, 507--532.

\bibitem[Debreu(1951)]{Debreu1951}
Debreu, G (1951), The coefficient of resource utilization. Econometrica
19, 273--92

\bibitem[Frege(1879)]{Frege1879}
Frege, G. (1879), Begriflsschriit, a formula language, modeled upon that of arithmetic, for pure thought 
\url{https://dn720006.ca.archive.org/0/items/gottlob-frege-begriffsschrift-english/Gottlob\%20Frege\%20-\%20Begriffsschrift\%20\%28English\%29_text.pdf}

\bibitem[Siekmann(2014)]{Siekmann2014}
Siekmann, J. (2014), Computational Logic, in Computational Logic edited by Jörg H. Siekmann, North Holland publications, pp: 15--30
\url{https://www.sciencedirect.com/science/chapter/handbook/abs/pii/B9780444516244500010}

\bibitem[Kirman(2010)]{Kirman2010}
Kirman, A. (2010), Complex Economics, Individual and Collective Rationality, London, Routledge.

\bibitem[Wilson and Kirman(2016)]{WilsonandKirman2016}
Wilson, D. and Kirman, A. (2016), Complexity and Evolution: Toward a New Synthesis for Economics, Cambridge, MA,  The MIT Press.

\bibitem[Hunter(2020)]{Hunter2007}
Hunter, J.D. (2007) Matplotlib: A 2D Graphics Environment, Computing in Science \& Engineering, vol. 9, no. 3, pp. 90--95.
 
\bibitem[Harris et. al.,(2020)]{Harrisetal2020}
Harris, C.R., Millman, K.J., van der Walt, S.J. et al. (2020), Array programming with NumPy. Nature 585, 357--362 (2020). DOI: 10.1038/s41586-020-2649-2.

\bibitem[Nosek et. al.,(2015)]{Noseketal2015}
Nosek BA, Alter G, Banks GC, Borsboom D, Bowman SD, Breckler SJ, Buck S, Chambers CD, Chin G, Christensen G, Contestabile M, Dafoe A, Eich E, Freese J, Glennerster R, Goroff D, Green DP, Hesse B, Humphreys M, Ishiyama J, Karlan D, Kraut A, Lupia A, Mabry P, Madon T, Malhotra N, Mayo-Wilson E, McNutt M, Miguel E, Paluck EL, Simonsohn U, Soderberg C, Spellman BA, Turitto J, VandenBos G, Vazire S, Wagenmakers EJ, Wilson R, Yarkoni T (2015). Promoting an open research culture. Science 348(6242):1422--1425. \href{https://doi.org/10.1126/science.aab2374}{https://doi.org/10.1126/science.aab2374}

\bibitem[Marx(1987)]{Marx1987}
Marx, K. (1987), Capital, Volume 1 2 3. International Publishers Company, Incorporated 
 
\bibitem[Schelling(1969)]{Schelling1969}
Schelling, T.C. (1969), Models of Segregation, The American Economic Review, Papers and Proceedings of the Eighty-first Annual Meeting of the American Economic Association,Vol. 59, No. 2, pp. 488--493.

\bibitem[Schelling(1978)]{Schelling1978}
Schelling, T.C. (1978), \emph{Micromotives and Macrobehavior (Fels Lectures on Public Policy Analysis)}, W. W. Norton \& Company.

\bibitem[Axelrod(1997)]{Axelrod1997}
Axelrod, R. (1997). The Dissemination of Culture: A Model with Local Convergence and Global Polarization: A Model with Local Convergence and Global Polarization. Journal of Conflict Resolution, 41(2), 203--226. \href{https://journals.sagepub.com/doi/10.1177/0022002797041002001}{https://journals.sagepub.com/doi/10.1177/0022002797041002001}

\bibitem[Lugo and Martínez-Mekler(2022)]{LugoMartinezMekler2022}
Lugo, I., and Martínez-Mekler, G. (2022). Theoretical study of the effect of ports in the formation of city systems. J. shipp. trd. 7, 16. \href{https://doi.org/10.1186/s41072-022-00117-6}{https://doi.org/10.1186/s41072-022-00117-6}

\bibitem[Mas-Colell(1995)]{MasColell1995}
Mas-Colell, A. (1995), Microeconomic Theory, Oxford University Press.

\bibitem[Wolfram(1983)]{Wolfram1983}
Wolfram, S. (1983). Statistical Mechanics of Cellular Automata. Rev. Mod. Phys. 55, 601-644.

\bibitem[de Souza Carvalho(2009)]{deSouzaCarvalho2009}
de Souza Carvalho, D. (2009), Cellular Automata Ordered by Entropy, Wolfram Demonstrations Project. 
\href{https://demonstrations.wolfram.com/cellularAutomataOrderedByEntropy/}
{Cellular Automata Ordered by Entropy}

\bibitem[Jost(2006)]{Jost2006}
Jost, L. (2006), Entropy and Diversity. Oikos, 2006, Vol. 113, No. 2 (May, 2006), pp. 363--375 Published by: Wiley on behalf of Nordic Society Oikos Stable.

\bibitem[Kerr and Goethel(2014)]{KerrandGoethel2014}
Kerr, L.A. and Goethel, D.R. (2014). Simulation Modeling as a Tool for Synthesis of Stock Identification Information, Editor(s): Steven X. Cadrin, Lisa A. Kerr, Stefano Mariani, Stock Identification Methods (Second Edition), Academic Press, pages 501--533, \href{doi:10.1016/B978-0-12-397003- 9.00021-7}{doi:10.1016/B978-0-12-397003- 9.00021-7}

\bibitem[Lugo et. al.,(2025)]{Lugoetal2025}
Lugo, I., Alatriste-Contreras, M.G. and Coutiño-Vázquez, B.G. (2025), Inclusive education via empathy propagation in schools of students with special education needs, 	arXiv:2511.16379 [cs.CY]. \href{https://arxiv.org/abs/2511.16379}{https://arxiv.org/abs/2511.16379} 

\bibitem[Molugaram and Rao(2017)]{MolugaramandRao2017}
Molugaram, K. and Rao, G. S. (2017), Random Variables, in Statistical Techniques for Transportation Engineering, Editor(s): Kumar Molugaram, G. Shanker Rao, Chapter 4, pp. 113--279, Butterworth-Heinemann..

\end{thebibliography}
\end{document}